# Soft Error Probability Estimation of Nano-scale Combinational Circuits

Ali Jockar
School of Electrical and Computer Engineering
Shiraz University
Shiraz, Iran
ajockar@shirazu.ac.ir

Mohsen Raji
School of Electrical and Computer Engineering
Shiraz University
Shiraz, Iran
mraji@shirazu.ac.ir

*Abstract*—As technology scales, nano-scale digital circuits face heightened susceptibility to single event upsets (SEUs) and transients (SETs) due to shrinking feature sizes and reduced operating voltages. While logical, electrical, and timing masking effects influence soft error probability (SEP), the combined impact of process variation (PV) and aging-induced degradation further complicates SEP estimation. Existing approaches often address PV or aging in isolation, or rely on computationally intensive methods like Monte Carlo simulations, limiting their practicality for large-scale circuit optimization. This paper introduces a novel framework for SEP analysis that holistically integrates PV and aging effects. We propose an enhanced electrical masking model and a statistical methodology to quantify soft error probability under process and aging variations. Experimental results demonstrate that the proposed approach achieves high accuracy while reducing computational overhead by approximately 2.5× compared to Monte Carlo-based methods. This work advances the design of reliable nano-scale circuits by enabling efficient, accurate SEP estimation in the presence of manufacturing variability and long-term transistor degradation.

I. INTRODUCTION

As technology continues to scale, feature sizes shrink, node capacitances decrease, and operating voltages drop. While these advancements improve performance and efficiency, they also raise significant concerns about the increased susceptibility of nano-scale digital circuits to single event upsets (SEUs) or soft errors [1][2]. These errors typically occur when a high-energy particle strikes a sensitive region within a transistor of a memory element (such as DRAM or SRAM cells), causing an unintended bit flip. Moreover, such particle strikes can also impact transistors in logic gates, generating transient electrical pulses known as Single Event Transients (SETs). The susceptibility of digital circuits to these errors is commonly measured using the Soft Error probability (SEP) metric.

Recent research and industrial observations suggest that, beyond SRAMs, latches, and flip-flops, combinational circuits are also becoming increasingly vulnerable to soft errors [3]-[5]. In fact, as reported in [6], the proportion of soft errors in combinational logic at the chip level has risen significantly—even at lower operating frequencies—as technology continues to advance.

One of the primary challenges in estimating the SEP of combinational circuits lies in three key masking effects [2]:
- Logical masking: If the logical values of a gate's inputs prevent the propagation of an SET, the transient pulse will be naturally filtered out.
- Electrical masking: The amplitude of the SET can diminish or completely disappear as it propagates through a sequence of logic gates.
- Timing masking: If an SET does not reach a memory element precisely within its required latching window (i.e., it does not meet setup and hold time constraints), the transient will have no observable effect.

As technology continues to advance, variations in cell geometries caused by manufacturing process, called as process variation (PV), become more significant, affecting circuit behavior at the nanoscale. Since a cell's resilience to transient faults induced by particle strikes is highly dependent on its physical dimensions [7], PV can significantly increase the impact of soft errors in combinational circuits. Furthermore, as circuits age, degradation mechanisms such as bias temperature instability (BTI) progressively impact transistor performance, making them more prone to soft errors over time. Given these challenges, an accurate assessment of the Soft Error Probability (SEP) in nano-scale circuit designs requires a comprehensive approach—one that accounts not only for PV but also for the gradual deterioration caused by aging.

Previous research on Soft Error Probability (SEP) in digital circuits have investigated PV or aging effects separately [8]-[9]. However, there are a few works in which the combined effect of both PV and aging is take into consideraion. The combined impact of process variations and aging presents a significant challenge in SEP estimation. In [10], a simulation-based approach was proposed to evaluate SEP under the influence of both factors. This appraoch consists of two key phases: fault generation and fault propagation. However, a major limitation of this method is the lack of an electrical masking model and an initial pulse width model, which affects the accuracy of the results. In contrast, [11] introduced a Monte Carlo-based analytical framework that considers both NBTI and process variations simultaneously. While Monte Carlo simulations offer high accuracy, their computational cost is significantly high, making them impractical for optimization algorithms.

In this paper, a novel appraoch is proposed for analyzing and improving the soft error reliability of digital circuits while simultaneously considering PV and aging effects. To achieve this, an electrical masking model is first introduced, followed by a statistical framework for soft error probability estimation, incorporating both PV and aging-induced degradations. Experimental results indicate that the proposed method, while maintaining high accuracy, achieves an approximately 2.5× speedup compared to Monte Carlo simulations.

## II. PROPOSED SEP ESTIMATION METHOD CONSIDERING PV AND AGING EFFECTS

This section presents the proposed for estimating the Soft Error Probability (SEP), incorporating both aging effects and process variations. In the following , the overl view of the proposed soft error probability estimation is provided and then, the three masking mechnaims are explained.

### A. Overal View of the proposed SEP estimation approach

In this section, we introduce our framework for estimating the Soft Error Rate (SEP) in the presence of process variations and aging effects. Figure 1 illustrates the flowchart of the proposed approach. As illustrated in Figure 1, the proposed framework consists of three main stages: (1) Initialization, (2) Transient Pulse Propagation, and (3) Soft Error Rate Calculation.

1. Initialization:This phase involves constructing a directed graph representation of the circuit, where nodes correspond to logic gates and edges represent signal interconnections. Subsequently, for each gate, we integrate process variation models, aging models, technology libraries, and the signal propagation probability (PP) computation. The statistical models used in this process are detailed in the subsequent sections.
2. Transient Pulse Propagatio: In this phase, all circuit gates are considered as potential candidates for transient pulse propagation. At each step, a target gate is selected as the impact site of an energetic particle strike, and all feasible paths from this gate to the circuit outputs are identified. A transient pulse is then selected based on an energy-dependent probability distribution from the initial pulse list, which contains pulses generated by high-energy particle strikes of varying energy levels. The proposed statistical model for initial pulse width is subsequently applied to the selected pulse. After identifying all accessible paths to the outputs, the electrical masking model is employed to propagate the final pulse width, and the probability of pulse propagation is computed. This process continues iteratively until all gates in the circuit have been evaluated.
3. Soft Error Probaility Calculation: In the final stage, we apply a maximum-likelihood estimation (MLE) approach with a Poisson-based probability model to aggregate the computed probabilities across all gates in the circuit. The final computed probability represents the SEP of the circuit.

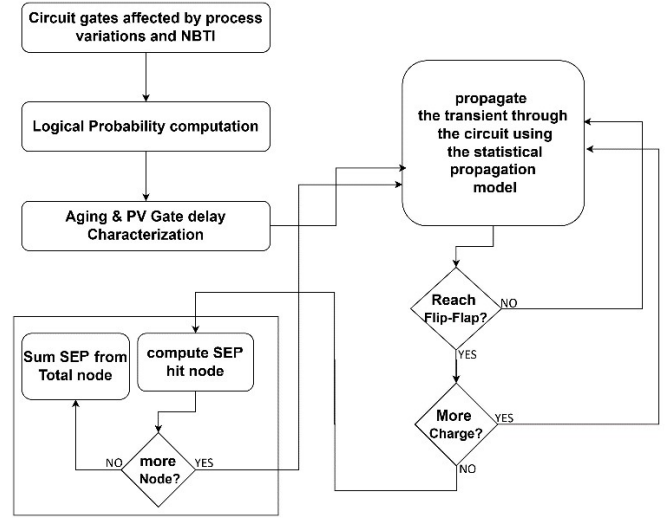

Figure 1- Flowchart of the Proposed Method for Estimating Soft Error Probability (SEP)

### B. Logical Masking

Logical masking plays a crucial role in transient pulse propagation within a circuit. In this section, we provide a detailed explanation of its impact on pulse propagation and describe the models utilized in the proposed appraoch.

**1. On-Path and Off-Path Signals**: when a high-energy particle strikes Gate A (which is considered a faulty candidate gate), a transient pulse with width W is generated at time t. The netlist, representing the propagation paths from the fault-injected gate to the reachable outputs, is categorized into two groups. The paths are extracted using a topological traversal approach to determine signal dependencies. On-path signals: Signals located along the propagation path from the fault site to the circuit outputs. The gates with at least one input originating from an on-path signal are referred to as on-path gates.Off-path signals: Signals that are not located along the primary propagation path. Gates in which all inputs originate from off-path signals are referred to as off-path gates [12][13].

**2. Transient Pulse Propagation Rules and Signal Propagation Probability (PP):** The probability of transient pulse propagation to the circuit output is influenced by the probability of off-path signals. In this work, the dependency is formulated using a probabilistic propagation model based on conditional probabilities between logic states as shown in Table 1. the propagation probability of on-path signals is determined by the probability of off-path signals.

Table 1: Transient Pulse Propagation Rules

| Gate | Rule |
|---|---|
| AND | $P_1(out) = \prod_{i=1}^{n} P_1(X_i)$ |
|  | $P_a(out) = \prod_{i=1}^{n} [P_1(X_i) + P_a(X_i)] - P_1(Out)$ |
| OR | $P_0(out) = \prod_{i=1}^{n} P_0(X_i)$ |
|  | $P_a(out) = \prod_{i=1}^{n} [P_0(X_i) + P_a(X_i)] - P_0(Out)$ |
| NOT | $P_0(out) = P_1(in)$ |
|  | $P_a(out) = P_{\bar{a}}(in)$ |
|  | $P_{\bar{a}}(out) = P_a(in)$ |
|  | $P_1(out) = P_0(in)$ |

In the proposed appraoch, we first extract all possible paths from fault injection sites to the circuit outputs. Then, we traverse all extracted paths and compute the probability of transient pulse propagation due to high-energy particle strikes at all outputs by applying the propagation rules in Table 1 for each gate. It is important to note that we assume the values of all signals, except for on-path signals (which propagate the transient pulse), remain constant. That is, no other signal contributes to transient pulse generation [12][14][15][16].

C. *Statistical Electrical Masking*

In [17], a mathematical framework is proposed for determining the width and amplitude of a transient pulse after passing through a logic gate, assuming a triangular or trapezoidal input pulse shape. When incorporating process variations into electrical masking, the model parameters become random variables with statistical distributions. Consequently, for different transition cases in the output pulse of a gate, the statistical distribution of the minimum and maximum output voltage, as well as the output pulse width, must be derived by assuming a normal distribution for the input pulse width and varying delay values. These distributions are derived from pre-characterized technology libraries and validated through SPICE-based circuit simulations to ensure accuracy across various operating conditions.

The output pulse width can be modeled as a function of the gate delay during pulse propagation and is computed based on Equations (1) and (2). Given that $PW_i$ and $PW_o$ are normally distributed random variables, the statistical distribution of $V_{Omin}$ and $V_{Omax}$ will also be normally distributed [18] and computed as:

$$V_{o_{min}} = (1 - PW_i / (t_f \times 1.25)) \times P(PW_i < t_f \times 1.25) + (0) \times (1 - P(PW_i < t_f \times 1.25)) \quad (1)$$

$$V_{o_{max}} = (PW_i / (t_r \times 1.25)) \times P(PW_i < t_r \times 1.25) + (V_{dd}) \times (1 - P(PW_i < t_r \times 1.25)) \quad (2)$$

Where:
- $V_{o_{min}}$ represents the maximum output voltage,
- $V_{o_{min}}$ represents the minimum output voltage.
- $PW_i$ is Input pulse width.
- $t_f$ falling transient pulse and $t_r$ rising transient pulse

These variables are random variables following a normal distribution.

Assuming a linear slope at the output, the minimum output voltage for a falling transient pulse (1→0) in an inverter is calculated using Equation (٦), where the random variables X and Y follow a normal distribution, as:

$$z = \frac{x}{y} \quad (٣)$$

This equation represents the probability that the input pulse width is less than 1.25 times the output fall time. Since both X and Y are normally distributed, their ratio follows a quotient distribution, which is a type of probability distribution that arises when dividing two random variables. If X and Y are independent normal variables, the resulting distribution becomes complex. Hinkley [18] proposed an approximation to simplify this quotient distribution into a normal distribution under certain conditions.

Similarly, for a rising transient pulse (0→1), the maximum output voltage can be computed using Equation (1), where X and Y follow normal distributions. This equation describes the probability that the input pulse width is less than 1.25 times the output rise time. Again, the statistical distribution follows a quotient distribution and can be approximated using a second-order Taylor expansion. The accuracy of this model is directly dependent on the precision of the input pulse width and the rise/fall times extracted from pre-characterized lookup tables. The proposed model is validated primarily for CMOS-based inverter circuits. While the appraoch can be extended to other logic gates, additional calibration may be required for FinFET or emerging transistor technologies.

The difference between two normal variables remains normally distributed. Assuming a fixed Vdd, the random variable X follows a normal distribution. The product of two normally distributed random variables remains normally distributed. The sum of two normal variables is also normally distributed. Initially, the statistical distribution of X is derived, followed by the computation of PWo based on the properties of normal distributions [17]. Using these approximations, we successfully derive the statistical distributions of $PW_o$, $d_1$, $PW_i$ as the as in Equation (4), leading to a robust statistical electrical masking model:

$$PW_o = (PW_i - d_1) + M \quad (4)$$

where $d_1$ and $M$ respectively indicate the delay of the first and preceding transient pulse edge.

Considering the impact of PV and BTI, the distribution of $PW_o$ is modeled as in Equation (5):

$$PW_o \sim Gauss(\mu_{PW_o}, \sigma^2_{PW_o})$$
$$\mu_{PW_o} = \mu_{PW_i} - \mu_{d_1} + \mu_M$$
$$\sigma^2_{PW_o} = \sigma^2_{sub} + \sigma^2_M - 2(\sigma^2_{sub}\sigma^2_M\sigma^2_{sub,M}) \quad (5)$$
$$\sigma^2_{sub} = \sigma^2_{PW_i} + \sigma^2_{d_2} - 2(\sigma^2_{PW_i}\sigma^2_{d_1}\sigma^2_{PW_i,d_1})$$

- $\rho_{PW_i,d_1}$ The correlation coefficient between the input pulse width and the initial propagation delay is considered **zero** in the proposed model.
- $\rho_{sub,M}$ The correlation coefficient between the random variable resulting from the difference (PW$_i$ & d$_1$) and the random variable resulting from the product (X and d$_2$) is considered. To simplify the analysis, these variables can be assumed to be independent.

Thus, we have derived the statistical distributions of $V_{o_{min}}$, $V_{o_{max}}$ and $PW_o$ develop a statistical electrical masking model.

### D. Statistical Timing Masking

As previously described, electrical masking influences the pulse width during propagation. Electrical masking not only attenuates the amplitude of the transient pulse but also reduces its effective width, thereby altering the likelihood of being latched by a flip-flop. Given this effect, the probability of a pulse being latched into a flip-flop can be computed using the following Equation (5) [14]:

$$LP = \frac{PW + LW}{Tclk} \quad (5)$$

where:
- $Tclk$ is the clock period of the circuit,
- $LW$ represents the latching window of the flip-flop,
- $PW$ denotes the width of the pulse reaching the output.

The accuracy of this probability model depends on the stability of $Tclk$ and variations in $LW$, which may fluctuate across different process technologies and operating conditions. Additionally, compared to computationally intensive Monte Carlo simulations, this analytical approach provides an efficient yet approximate estimation of pulse latching probability in synchronous circuits.

### E. SEP Estimation Computaiton

The probability of a soft error occurrence is determined as:

$$SEP = PP \times LP \quad (6)$$

where *lp* represents the probability of the pulse being latched at the output, and *pp* denotes the probability of signal propagation [14].

During the propagation of a transient pulse from the impact site, the soft error probability Epp is computed using a statistical framework based on maximum likelihood estimation (MLE), ensuring that the highest probable soft error occurrence is considered across all feasible paths leading to the circuit outputs.

The final soft error probability follows a Poisson-like statistical distribution, capturing the stochastic effects introduced by process variations and aging mechanisms. Compared to conventional Monte Carlo-based simulations, this analytical approach provides a computationally efficient estimation of soft error probability while maintaining a high level of accuracy.

## III. EXPERIMENTAL RESULTS

In order to show the efficacy of the proposed SEP estimation method, we conduct a set of experiments comparing the proposed method with Monte Carlo simulations as the baseline method for SEP estimation of combinational circuits. In the following, we first explain the experiment setup and then, the accuracy and computational efficiency of the proposed method are compared against Monte Carlo simulations in terms of error probability estimation precision and execution time.

### A. Experiment Setup

All simulations were implemented in C++ and executed in a Windows 10 environment. The experiments were conducted on a system equipped with an Intel Core i5 processor (3.4 GHz), 16 GB of RAM, and a 512 GB SSD with sequential read and write speeds of 1200 MB/s and 1800 MB/s, respectively. Additionally, an Intel UHD Graphics 620 GPU was utilized for data visualization and acceleration of computational tasks where applicable. The benchmark circuits employed in this study were selected from the ISCAS'85 benchmark suite. Table 2 shows the benchmark circuits and their relevant information.

### B. Comparison of the Proposed SEP Estimation approach with Monte-Carlo: Accuracy

Aging effects were evaluated for 3, 6, and 9 years, while process variations were considered with variance levels of 5%, 10%, and 20%. Tables 3 to 6 present the results obtained from Monte Carlo simulations and the proposed method for different process variation reation and different years of aging. On average, the proposed method exhibits a 4% to 8% deviation from Monte Carlo results. This discrepancy arises from the use of statistical models instead of iterative Monte Carlo sampling. However, this margin of error is acceptable considering the significant improvement in execution time achieved by the proposed method.

| Table 2 :Circuit & gateinfo | | | | |
|---|---|---|---|---|
| Circuit | #gate | #PI | #PO | LvMax |
| c432 | 120 | 36 | 7 | 30 |
| c499 | 162 | 41 | 32 | 28 |
| c880 | 320 | 60 | 26 | 33 |
| c1355 | 506 | 41 | 32 | 30 |
| c1908 | 603 | 33 | 25 | 39 |
| c2670 | 872 | 233 | 140 | 38 |
| c3540 | 1179 | 50 | 22 | 52 |
| c5315 | 1726 | 178 | 123 | 41 |
| c6288 | 2384 | 32 | 32 | 122 |

**Table 3 : Soft Error Probability (10% Variations & 3-Year Aging Impact)**

| Circuit | Monte-Carlo | | Proposed method | | The differences % | |
|---|---|---|---|---|---|---|
| | μ | σ | μ | σ | μ | σ |
| c432 | 1.18 | 0.11 | 1.22 | 0.12 | 3.28 | 8.33 |
| c499 | 1.76 | 0.96 | 1.82 | 1.02 | 3.30 | 5.88 |
| c880 | 6.67 | 0.23 | 6.84 | 0.25 | 2.49 | 8.00 |
| c1355 | 5.36 | 0.18 | 5.58 | 0.20 | 3.94 | 10.00 |
| c1908 | 2.43 | 0.85 | 2.48 | 0.90 | 2.02 | 5.56 |
| c2670 | 4.49 | 0.75 | 4.63 | 0.80 | 3.02 | 6.25 |
| c3540 | 0.66 | 0.35 | 0.67 | 0.39 | 1.49 | 10.26 |
| c5315 | 0.60 | 0.27 | 0.62 | 0.29 | 3.23 | 6.90 |
| c6288 | 8.31 | 0.52 | 8.57 | 0.56 | 3.03 | 7.14 |
| c7552 | 0.69 | 0.27 | 0.71 | 0.29 | 2.82 | 6.90 |

**Table 4 : Soft Error Probability (20% Variations & 3-Year Aging Impact)**

| Circuit | Monte-Carlo | | Proposed method | | The differences % | |
|---|---|---|---|---|---|---|
| | μ | σ | μ | σ | μ | σ |
| c432 | 1.18 | 0.22 | 1.24 | 0.25 | 4.84 | 12.00 |
| c499 | 1.76 | 0.19 | 1.79 | 0.22 | 1.68 | 13.64 |
| c880 | 6.67 | 0.47 | 6.85 | 0.53 | 2.63 | 11.32 |
| c1355 | 5.36 | 0.35 | 5.47 | 0.4 | 2.01 | 12.50 |
| c1908 | 2.43 | 0.17 | 2.53 | 0.19 | 3.95 | 10.53 |
| c2670 | 4.49 | 0.14 | 4.63 | 0.15 | 3.02 | 6.67 |
| c3540 | 0.66 | 0.7 | 0.68 | 0.77 | 2.94 | 9.09 |
| c5315 | 0.6 | 0.54 | 0.61 | 0.62 | 1.64 | 12.90 |
| c6288 | 8.2 | 0.62 | 8.44 | 0.69 | 2.84 | 10.14 |
| c7552 | 0.7 | 0.53 | 0.73 | 0.59 | 4.11 | 10.17 |

**Table 5 : Soft Error Probability (10% Variations & 9-Year Aging Impact)**

| Circuit | Monte-Carlo | | Proposed method | | The differences % | |
|---|---|---|---|---|---|---|
| | μ | σ | μ | σ | μ | σ |
| c432 | 1.29 | 0.11 | 1.32 | 0.12 | 2.27 | 8.33 |
| c499 | 1.92 | 0.1 | 2.03 | 0.11 | 5.42 | 9.09 |
| c880 | 7.29 | 0.25 | 7.69 | 0.28 | 5.20 | 10.71 |
| c1355 | 5.86 | 0.19 | 6.12 | 0.21 | 4.25 | 9.52 |
| c1908 | 2.66 | 0.92 | 2.8 | 1 | 5.00 | 8.00 |
| c2670 | 4.91 | 0.8 | 5.15 | 0.9 | 4.66 | 11.11 |
| c3540 | 0.73 | 0.38 | 0.77 | 0.43 | 5.19 | 11.63 |
| c5315 | 0.65 | 0.29 | 0.68 | 0.32 | 4.41 | 9.38 |
| c6288 | 8.57 | 0.97 | 9.01 | 1.06 | 4.88 | 8.49 |
| c7552 | 0.76 | 0.29 | 0.80 | 0.32 | 5.00 | 9.38 |

**Table 6 : Soft Error Probability (20% Variations & 9-Year Aging Impact)**

| Circuit | Monte-Carlo | | Proposed method | | The differences % | |
|---|---|---|---|---|---|---|
| | μ | σ | μ | σ | μ | σ |
| c432 | 1.29 | 0.23 | 1.35 | 0.26 | 4.44 | 11.54 |
| c499 | 1.92 | 0.20 | 2.02 | 0.23 | 4.95 | 13.04 |
| c880 | 7.28 | 0.51 | 7.71 | 0.58 | 5.58 | 12.07 |
| c1355 | 5.86 | 0.39 | 6.19 | 0.45 | 5.33 | 13.33 |
| c1908 | 2.66 | 0.18 | 2.81 | 0.20 | 5.34 | 10.00 |
| c2670 | 4.91 | 0.16 | 5.15 | 0.18 | 4.66 | 11.11 |
| c3540 | 0.73 | 0.77 | 0.77 | 0.89 | 5.19 | 13.48 |
| c5315 | 0.65 | 0.59 | 0.68 | 0.67 | 4.41 | 11.94 |
| c6288 | 9.40 | 0.76 | 9.95 | 0.86 | 5.53 | 11.63 |
| c7552 | 0.76 | 0.58 | 0.80 | 0.67 | 5.00 | 13.43 |

## C. Comparison of the Proposed SEP Estimation approach with Monte-Carlo: Runtime

Figure 2 compares the execution time of soft error rate (SEP) estimation using the proposed method and the Monte Carlo method. The execution time, measured in milliseconds, is presented on a logarithmic scale. The blue bars correspond to the Monte Carlo method, while the orange bars represent the proposed method. As shown in Figure 2, the proposed method achieves a remarkable reduction in execution time compared to the Monte Carlo approach. Specifically, the proposed method is approximately 2.5 times faster than the Monte Carlo method. This significant reduction in execution time makes the proposed method highly suitable for integration into optimization algorithms, enabling rapid SEP recalculations after each circuit improvement. In contrast, the Monte Carlo method's excessively long execution time renders it impractical for use in such iterative optimization processes.

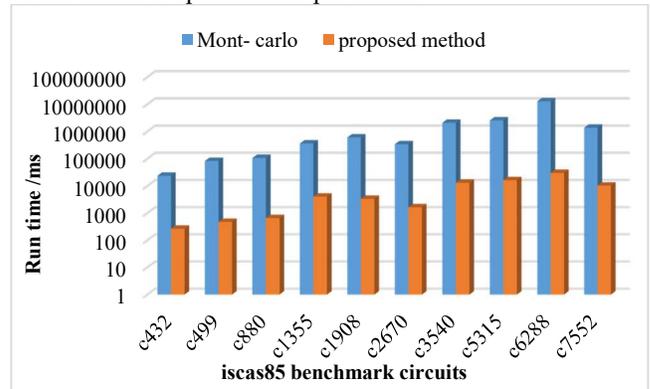

Figure 2-The execution time for soft error probability estimation using the statistical approach is compared with the Monte Carlo method.

## IV. CONCLUSION

Soft errors have evolved into a significant concern in the design of nanoscale digital circuits [1], [2]. With advancements in manufacturing technology and the continuous scaling of transistor dimensions, transient faults caused by high-energy particle strikes have led to a notable decline in circuit reliability [3], [4]. The reduction in transistor size, combined with two critical factors—process variations and aging—has become a primary challenge in accurately estimating the soft error rate (SEP) resulting from high-energy particle strikes at the gate level [5], [6]. Process variations, such as variations in threshold voltage and channel length, directly impact the critical charge required to trigger a soft error. Similarly, aging mechanisms like Negative Bias Temperature Instability (NBTI) and Hot Carrier Injection (HCI) degrade transistor performance over time, making circuits more susceptible to soft errors . In the first part of this thesis, we introduce a statistical approach to compute the soft error probability, taking into account both process variations and aging effects simultaneously.